\documentclass[fleqn,usenatbib]{mnras}

\usepackage{amsmath}
\usepackage{times,txfonts}
\usepackage[T1]{fontenc}
\usepackage{ae,aecompl}

\usepackage{graphicx}	
\usepackage{amssymb}	
\usepackage{bm}
\usepackage{threeparttable}
\usepackage{multicol}
\usepackage{ulem}
\usepackage{booktabs}
\usepackage{comment}
\usepackage{multirow}
\usepackage[title]{appendix}
\DeclareMathAlphabet{\mathcal}{OMS}{cmsy}{m}{n}

\title[RCS in Flaring Magnetospheres]{Spectral Modification of Magnetar Flares by Resonant Cyclotron Scattering}

\author[Yamasaki, Lyubarsky, Granot \& G{\"o}{\u{g}}{\"u}{\textcommabelow s}]
{Shotaro Yamasaki\thanks{E-mail: shotaro.s.yamasaki@gmail.com}$^{1,2,3}$, Yuri Lyubarsky$^{4}$, Jonathan Granot$^{5,6}$, and  Ersin G{\"o}{\u{g}}{\"u}{\textcommabelow s}$^{7}$
\\
$^{1}$Racah Institute of Physics, The Hebrew University of Jerusalem, Jerusalem 91904, Israel\\
$^{2}$The Raymond and Beverly Sackler School of Physics and Astronomy, Tel Aviv University, Tel Aviv 69978, Israel\\
$^{3}$Institute for Cosmic Ray Research, The University of Tokyo,
5-1-5, Kashiwanoha, Kashiwa, Chiba, 277-8582, Japan\\
$^{4}$Department of Physics, Ben Gurion University of the Negev, P.O. Box 653, Beer-sheba 84105, Israel\\
$^{5}$Department of Natural Sciences, The Open University of Israel, P.O Box 808, Ra'anana 43537, Israel\\
$^{6}$Department of Physics, The George Washington University, Washington, DC 20052, USA\\
$^{7}$Faculty of Engineering and Natural Sciences, Sabanc{\i} University, Orhanl{\i} Tuzla 34956, {\.{I}}stanbul, Turkey
}

\date{Accepted XXX. Received YYY; in original form ZZZ}

\pubyear{2020}

\begin{document}
\label{firstpage}
\pagerange{\pageref{firstpage}--\pageref{lastpage}}
\maketitle

\begin{abstract}
Spectral modification of energetic magnetar flares by resonant cyclotron scattering (RCS) is considered. During energetic flares, photons emitted from the magnetically-trapped fireball near the stellar surface should resonantly interact with magnetospheric electrons or positrons. We show by a simple thought experiment that such scattering particles are expected to move at mildly relativistic speeds along closed magnetic field lines, which would slightly shift the incident photon energy due to the Doppler effect. 
We develop a toy model for the spectral distortion by a single RCS that incorporates both a realistic seed photon spectrum from the trapped fireball and the velocity field of particles, which is unique to the flaring magnetosphere. 
We show that our spectral model can be effectively characterized by a single parameter; the effective temperature of the fireball, which enables us to fit observed spectra with low computational cost. We demonstrate that our single scattering model is in remarkable agreement with {\it Swift}/BAT data of intermediate flares from SGR 1900+14, corresponding to effective fireball temperatures of $T_{\rm eff}=6$--$7$ keV, whereas {\it BeppoSAX}/GRBM data of giant flares from the same source may need more elaborate models including the effect of multiple scatterings.
Nevertheless, since there is no standard physically-motivated model for magnetar flare spectra, our model could be a useful tool to study magnetar bursts, shedding light on the hidden properties of the flaring magnetosphere. 
\end{abstract}

\begin{keywords}
stars: flare, magnetars, neutron -- X-rays: bursts -- scattering.
\end{keywords}



\section{Introduction}
\label{s: Introduction}

Magnetars \citep{DT1992,Paczynski1992}, a class of strongly magnetized neutron stars (with surface dipole field strength $B_p\sim10^{14}$--$10^{15}\;$G), are known to exhibit flaring activities over a wide range of luminosities ($L\sim10^{38}$--$10^{47}\:{\rm erg\:s^{-1}}$), most of which is unleashed in X-rays and soft gamma-rays. They are phenomenologically classified into ``giant flares" ($10^{44}$--$10^{47}$ erg $\rm s^{-1}$) emitted in several minutes, ``intermediate flares" ($10^{41}$--$10^{43}$ erg $\rm s^{-1}$) and ``short bursts"
($10^{38}$--$10^{41}$ erg ${\rm s^{-1}}$) with duration ranging from a few millisecond to a few second (for recent reviews, see \citealt{Kaspi2017,Enoto2019}).
Remarkably, the fluence of magnetar flares (from short bursts to giant flares) broadly follow a single power-law distribution $N\propto {F}^{-\alpha}$ with an index of $\alpha\sim1{\rm -}2$ (e.g., \citealt{Cheng1996,Gogus2001,Woods2006,Nakagawa2007,Collazzi15,Lin20}). 
These bursts are believed to be generated by a sudden release of magnetic energy, which would result in the formation of a hot electron/positron plasma.
With some exceptions (e.g., the initial short hard spike of giant flares), the fireball is confined near the stellar surface by the strong magnetic pressure, thereby forming an optically-thick bubble called ``trapped fireball'' \citep{TD1995}. The trapped fireball gradually cools by losing its energy through the radiation from its photosphere and occasionally manifests itself as a soft extended tail which shows high-amplitude pulsations over $1$-$100\;$s at the same spin period of an underlying neutron star occasionally manifests itself as a soft and minutes long burst tail which shows high-amplitude periodic modulations ($1$-$100\;$s) at the same spin period of an underlying neutron star \citep{TD1995,TD1996,Feroci2001}.
While the above process is likely to operate in energetic flares such as giant flares and intermediate flares, whether the fireball successfully forms in the lesser flares (i.e., short bursts) remains unclear (e.g., \citealt{Watts2010,Kaspi2017}). This is partly because the observationally inferred sizes of the emission regions for lesser flares are so small (e.g., \citealt{Gogus2000}) that they are indistinguishable from the inferred sizes of the hot spots on the stellar surface \citep{Yamasaki2019}.

Early considerations of photon-energy-dependent radiative transport inside the trapped fireball predicted the observed burst emission spectrum (the photon flux per unit energy) to appear almost flat at the Rayleigh-Jeans region 
and to remain the same as a blackbody at the Wien region (\citealt{Lyubarsky2002}, see also \citealt{Ulmer1994,Miller1995}).
This is due to the energy dependence of the opacity for photons in the extraordinary polarization mode expected under the presence of strong magnetic fields. This allows the lower energy photons to escape from deeper parts of the fireball, and thus the radiation at low energies emerges as a superposition of blackbodies, shaping the flat spectrum (see \S \ref{sss:Seed Photon Spectrum}).
Later, the observed flaring spectra at soft X-ray energies appeared in good agreement with the model, whereas the model significantly underpredicts the observed spectra at hard X-ray energies (see e.g., Figure 9 of \citealt{Olive2004,Israel2008}), and the discrepancy remains unsolved for more than a decade. 

The resonant cyclotron scattering (RCS) may well be a plausible process that can
help explain the observed spectra
of energetic magnetar flares.
Magnetars emit mostly in the X-ray band so that close to their surface, the cyclotron frequency well exceeds the radiation frequency. But at the distance 5-10 neutron star radii, the radiation passes the cyclotron resonance layer. At the resonance, the effective cross section exceeds the classical Thomson value by at least a few orders of magnitude. The magnetar magnetosphere is filled with electron-positron plasma both during flares and in the persistent state \citep{TLK2002,Beloborodov2007,Beloborodov2013}; one can easily see that the cyclotron optical depth is large. Therefore all the outgoing radiation is reprocessed in the cyclotron resonance layer; one has to take this into account when analyzing the observed properties of magnetar emission. The full problem of the radiation transfer through the cyclotron resonance layer is extremely complicated. Even in the two-level approximation, the mechanisms of the radiation escape in the line wing has been figured out only recently \citep{Garasev2008,Garasev2011,Garasev2016}.  In any case, one has to take into account that the magnetospheric plasma is by no means at rest. 
Since the scattering particles are expected to move along
the magnetic field lines,
the energy of scattered photons would shift due to the relativistic Doppler effect. Hence the velocity distribution of the scattering particles is of profound importance in the RCS. 

Historically, the RCS has been primarily studied in the context of modeling the spectra of the quiescent magnetar emission (e.g., \citealt{TLK2002,Baring2007,LG2006,FT2007,Beloborodov2007,Guver2007,Nobili2008,Rea2008,Zane2011,Beloborodov2013,Wadiasingh2018}), which is less luminous (typically $L\lesssim10^{35}\ {\rm erg\ s^{-1}}$) compared to magnetar flares by 
$\sim3-12$ orders of magnitude. However, a detailed model of RCS during the bursting phase has yet to be developed.
In the quiescent state,
magnetospheric particles with relativistic and/or ultra-relativistic velocity may be present \citep{Beloborodov2007}. In contrast,
during the flare, tremendous resonance radiation force makes the plasma move mildly relativistically, the plasma bulk velocity at any point being determined by the condition that the radiation is directed, in the plasma comoving frame, perpendicularly to the local magnetic field \citep{Beloborodov2013}. This very fact produces distortions in the outgoing spectrum because the rescattered photons are Doppler shifted. 

Here we develop, as the first step towards more elaborate models of the radiation reprocessing in the magnetar magnetosphere, a toy model assuming that photons escape after one scattering in the resonance layer. Our aim is to demonstrate that the Doppler shift due to scattering on the bulk motions of the magnetospheric plasma could lead to formation of hard tails in thermal spectra. We believe that this qualitative result is robust and not very sensitive to details of the frequency redistribution within the cyclotron line. Therefore we ignore many nuances of the resonance scattering process, such as the angular and polarization dependence of the resonant cross section, spin flip processes, relativistic corrections, transitions to higher Landau levels etc. Since our spectral model can be effectively described by a single parameter -- the effective temperature of the fireball, this greatly reduces the parameter space
and allows us to fit the observed spectra with low computational cost. Our model is expected to be applicable to energetic flares, such as intermediate and giant flares, for which fireballs are likely present due to their high energy dissipation rates.

Recently, one of the most prolific transient magnetar, SGR J1935+2154 went into an intense bursting episode, during which hundreds of energetics bursts, including quite a number of intermediate flares, were recorded \citep{Palmer20,nicergbm_paper}. Interestingly, one of those X-ray bursts observed with numerous orbiting telescopes \citep{integral_paper,hxmt_paper,konus_paper,agile_paper} was temporarily associated with the extremely bright millisecond radio burst \citep{Chime,Stare2}, which is reminiscent of cosmological fast radio bursts (FRBs). It is intriguing that the X-ray burst associated with FRB-like radio burst has an unusually hard spectrum compared to other X-ray bursts with comparable (or even higher) fluence and the FRB-like counterpart was seen only in one of many X-ray bursts. Even this fact alone makes a better understanding of energetic magnetar flares essential.

This paper is organized as follows.  In \S \ref{s:Formalism}, we introduce the characteristic velocity field of particles in the flaring magnetosphere.  We present our toy model and describe in detail how to implement the simulation in \S \ref{s:Model}, followed by the simulation results in \S \ref{s:Simulation}.
Our model will be applied to the observed burst spectra from SGR 1900$+$14 in \S \ref{s:Application}. A final discussion appears in \S \ref{s:Discussion}. We summarize our findings in \S \ref{s:Conclusions}.

\section{Equilibrium Particle Velocity Field in Flaring Magnetospheres}
\label{s:Formalism}

In the magnetar magnetospheres, an electron can be effectively treated being restricted to move along the magnetic field line (as beads threaded on a wire) in its lowest Landau level since it loses its gyro-momentum through the fast cyclotron cooling, which leads to the rapid decay of excited quantum states\footnote{We consider the simplest transition from the first Landau state to the ground state. However, this is the case only if the incident photon's angle in the electron's rest frame is very small \citep{Gonthier2000} and otherwise higher intermediate states and final states may be accessible \citep{Herold1979,Bussard1986,Daugherty1986}. Nevertheless, such uncertainties must be 
sub-dominant relative to the strongest assumption of a single scattering (see \S \ref{ss:Scattering}).
}.
In these conditions, a sequence of two independent processes; absorption and re-emission can be regarded as a single scattering with the  non-relativistic  classical resonant cross section
(e.g., \citealt{Canuto1971})
\begin{equation}
\label{eq:sigma_res}
    \sigma_{\rm res}(\omega)=\pi^2r_e c\,\delta(\omega-\omega_B)(1+\cos^2\theta_i),
\end{equation}
where $\omega_B\equiv eB/(m_e c)$ is the cyclotron frequency in the electron rest frame (ERF) with $B$ being the local magnetic field strength, $r_e\equiv e^2/(m_ec^2)$ the classical electron radius and $\theta_i$ the angle of incoming
photon measured in the ERF with respect to the particle momentum. As the electron's momentum is parallel to the local magnetic field direction, $B$ conserves between ERF and observer frame (OF). 

\begin{figure*}
 \includegraphics[scale=0.56]{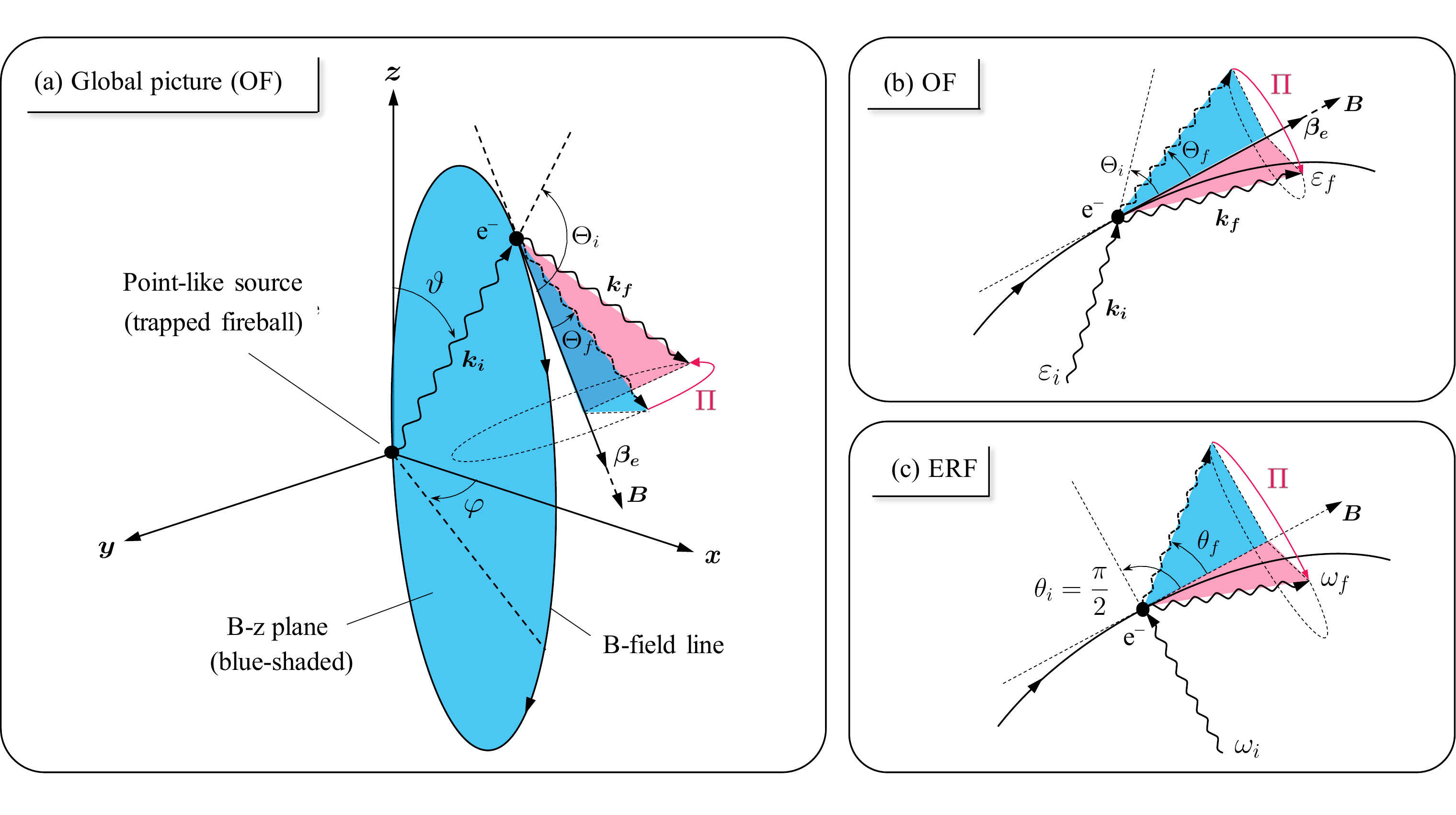}
 \caption{Scattering geometry in (a)--(b) the observer frame (OF) and (c) the electron's rest frame (ERF); all relevant angles and energies are indicated. The dipole magnetic moment is taken to be parallel to the z-axis. 
}
 \label{fig:schematic}
\end{figure*}

Let us consider the radiation from a magnetically-confined fireball formed during the flare. Here the fireball is approximated as a point-like source of isotropic emission located at the center of the neutron star (see also \S \ref{ss:Scattering} for details). 
These photons are efficiently scattered by magnetospheric particles if the photon energy in the ERF satisfies a resonance condition:
\begin{equation}
\label{eq:resonance condition}
\omega_i=\gamma_e\epsilon_i(1-\beta_e\cos\Theta_i)= \hbar\omega_B
\end{equation}
where $\gamma_e$ is the Lorentz factor of a scattering particle, $\beta_e$ the particle velocity in units of $c$, $\Theta_i$ the angle of incoming photon measured in the OF with respect to the particle momentum (see Table \ref{tab: description of key quantities} and Figure \ref{fig:schematic}). 
In general, the location of the resonance layer depends on the magnetic field structure and the particle velocity field. 
For a dipole magnetic field geometry, the cyclotron energy of an electron at a given distance $r$ from the stellar center can be written as 
\begin{equation}
\label{eq:cyclotron energy}
\hbar\omega_B\sim1.1\ B_{p,14}\,\left(r/R_{\rm NS}\right)^{-3}\ {\rm MeV},
\end{equation}
where $B_{p,14}=B_p/(10^{14}\,{\rm G})$ is the polar magnetic field strength and $R_{\rm NS}$ the stellar radius for
which we assume a typical value of $10$ km. The resonance layer at which the resonance condition (eq.[\ref{eq:resonance condition}]) is met for $\epsilon_i\lesssim10$ keV photons locates at $\gtrsim$ several stellar radii if the factor of $\gamma_e(1-\beta_e\cos\Theta_i)$ is neglected. The assumption of the point-like fireball therefore seems valid except for the most energetic giant flares that might generate a trapped fireball with its size comparable to the altitude of resonance layer.

\begin{table}
\centering
\caption{Description of important quantities that control RCS.}
\label{tab: description of key quantities}
\begin{tabular}{ll}
 \cmidrule[0.8pt](r){1-2}
$\epsilon_{i(f)} \cdots$    & Initial (final) photon energy in the OF  \\
$\omega_{i(f)}\cdots$     & Initial (final) photon energy in the ERF \\
$\Theta_{i(f)} \cdots$    & Initial (final) photon angle to the particle momentum \\ & (local magnetic field) in the OF\\
$\theta_{i(f)} \cdots$    & Initial (final) photon angle to the particle momentum \\ & (local magnetic field) in the ERF\\
$\bm{\hat{k}}_{\bm{i(f)}} \cdots$    & Unit vector of initial (final) photon momentum \\&in the OF\\
$\vartheta \cdots$    & Colatitude (polar angle) of initial photon\\
$\varphi \cdots$    & Azimuthal angle of initial photon about $\bm{\hat{z}}$\\
$\Pi \cdots$    & Azimuthal angle of scattered photon about $\bm{\hat{B}}$\\&in the ERF ($\Pi=0$ in $\bm{\hat{B}}$-$\bm{\hat{z}}$ plane)\\
\cmidrule[0.8pt](r){1-2}
\end{tabular}
\end{table}

During the magnetar flare in the luminosity range of interest $L\gtrsim10^{40}\:{\rm erg\:s^{-1}}$, magnetospheric particles should feel a strong radiation drag force. This is in stark contrast to the case of persistent emissions from magnetars in the quiescent state, which is much less powerful $L\lesssim10^{35}\:{\rm erg\:s^{-1}}$.
We consider the following thought experiment on the motion of an electron bathed in such a strong radiation field. The incident photon angle with respect to the local magnetic field in the ERF is related to the OF angle by Lorentz transformation as
\begin{equation}
\label{eq:Doppler trans 1}
\cos\theta_i=\frac{\cos\Theta_i-\beta_e}{1-\beta_e\cos\Theta_i}.
\end{equation}
Thus, an electron moving parallel to the magnetic field with speed $\beta_e\sim0$ (i.e., the right-hand side of eq. [\ref{eq:Doppler trans 1}] is positive) is pushed forward by the radiation force ($\theta_i<\pi/2$), which leads to acceleration. Conversely, once the electron attains a relativistic velocity $\beta_e\sim1$ (i.e., the right-handed side of eq. [\ref{eq:Doppler trans 1}] is negative), it is pushed back by the radiation force ($\theta_i>\pi/2$), which leads to deceleration. Therefore, even if the electron does not ``see" the photon with right angle, it would be immediately accelerated or decelerated to reach the equilibrium state, where the radiation force is directed {\it perpendicularly} to the magnetic field ($\theta_i=\pi/2$)\footnote{A qualitatively similar arguments are made by \citet{Beloborodov2013} in the context of quiescent emission from magnetars.}.  Since the timescale for this regulation is negligibly short (see Appendix \ref{s: tau_relax}), we can reasonably assume the above condition in the ERF, which allows us to uniquely determine the particle velocity in resonance with an incoming photon with $\Theta_i$ as
\begin{equation}
\label{eq:velocity distribution}
\beta_e=\cos\Theta_i; \quad\quad \gamma_e=1/\sin\Theta_i.
\end{equation}
A direct consequence of this velocity field might be the accumulation of decelerated plasma near the highest point of each closed magnetic field line (e.g., \citealt{Beloborodov2013}). Tellingly, these particles might annihilate and emit $m_e c^2\sim511$ keV lines although it is highly dependent on the local plasma density and thus out of the scope of this work. 

The RCS process does not change the photon energy in the ERF; $\omega_i=\omega_f$, since the majority of $\lesssim{\cal O}(10\ {\rm keV})$ photons satisfies $\epsilon_i\ll m_e c^2/\gamma_e$ with $\gamma_e\sim{\cal O}(1)$ and thus the electron recoil is negligible. The scattered photon energy in the OF, $\epsilon_f$, is related to the emission angle in the ERF, $\theta_f$, through the Lorentz transformation of $\omega_f$,
\begin{equation}
\label{eq: epsilon_f}
\epsilon_f=\gamma_e\omega_f(1+\beta_e\cos\theta_f).
\end{equation}
The photon emission angle in the ERF, $\theta_f$, is in a random direction (for a detailed implementation, see \S \ref{sss:Emission Angle}). Finally, the photon emission angle in the OF, $\Theta_f$, is given by 
\begin{equation}
\label{eq:Doppler trans 2}
\cos\Theta_f=\frac{\cos\theta_f+\beta_e}{1+\beta_e\cos\theta_f},
\end{equation}
which is equivalent to equation \eqref{eq:Doppler trans 1} via the inverse Lorentz transformation.

\section{A Toy Model}
\label{s:Model}

We aim to obtain the reprocessed spectrum of the fireball emission by injecting the seed photons by three dimensional Monte Carlo method. Below, detailed implementation and key assumptions are presented.

\subsection{Model Geometry}
In order to track a single photon trajectory, we adopt spherical coordinate $r$, $\theta$, $\phi$ centered on the star with $z$-axis aligned with magnetic pole.
When the twist of magnetic fields relative to magnetic poles is moderate, poloidal magnetic fields are well approximated as dipole \citep{TLK2002}:
\begin{equation}
\label{eq:B-field}
\bm{\hat{B}}=\frac{2\cos\theta\,\bm{\hat{r}}+\sin\theta\,\bm{\hat{\theta}}}{\sqrt{3\cos^2\theta+1}},
\end{equation}
where $\bm{\hat{r}}$ and $\bm{\hat{\theta}}$ denote basis vectors for polar coordinate. Since the scattering process is solely dependent on the configuration of magnetic fields that uniquely determine the particle velocity field (eq. [\ref{eq:velocity distribution}] in \S \ref{s:Formalism}), no assumption is made on the magnetic field strength. 

Although the emission from the point-like fireball is isotropic, the emission observed along a given line of sight might be modulated by the rotation of the star (which will be found not to be the case in \S \ref{s:Simulation}). As a first step to obtain the spectra averaged over the entire rotational phase, we consider an aligned rotator in which magnetic axis is parallel to
the spin axis ($\bm{\hat{\Omega}}=\bm{\hat{\mu}}_B=\bm{\hat{z}}$).

\subsection{Scattering}
\label{ss:Scattering}
\subsubsection{Seed Photon Spectrum}
\label{sss:Seed Photon Spectrum}
We adopt a primary X-ray photon energy spectrum of emission from the trapped fireball proposed by \citet{Lyubarsky2002}, who considered the detailed radiation transfer in the fireball under strong magnetic fields.
The spectral formation inside the fireball is strongly affected by the presence of the two polarization modes with different scattering cross sections: the ordinary mode (O-mode: polarized in the $\hat{\bm k}$-$\hat{\bm B}$ plane) and extraordinary mode (E-mode: perpendicularly polarized to the $\hat{\bm k}$-$\hat{\bm B}$ plane). Since the scattering of the E-mode photons is significantly suppressed by a factor of $\sigma_{\rm E}/\sigma_{\rm O}\sim(\epsilon/\hbar\omega_B)^2\sim10^{-4}(\epsilon/10{\: \rm keV})^2(B/10^{14}{\:\rm G})^{-2}$ \citep{Meszaros1992}, the photosphere of the E-mode photons lies far below that of O-mode photons. This allows the observer to see deeper into the fireball at lower energies, where each layer of E-mode photosphere radiates a Planckian spectrum. 
As a result, the emerging spectrum from the trapped fireball has a {\it non-Planckian} form:
\begin{equation}
\label{eq:fireball spectrum}
N(\epsilon)\propto\epsilon^2\left\{\exp\left[\frac{\epsilon^2}{T_{\rm eff}\sqrt{\epsilon^2+(3\pi^2/5)T_{\rm eff}^2}}\right]-1\right\}^{-1},
\end{equation}
where $T_{\rm eff}$ is the effective (bolometric) temperature of the fireball. 
The overall spectrum is characterized by a plateau at Rayleigh-Jeans region due to the energy dependence of the E-mode opacity ($\propto\epsilon^2$), which is in striking contrast to the commonly assumed Planckian spectrum. 
Before examining the broadband seed photon spectrum (eq. [\ref{eq:fireball spectrum}]), a mono-energetic spectrum with $N(\epsilon)\propto\delta(\epsilon-T_{\rm eff})$ will be explored in order to see qualitatively how our model redistributes
the photon energy.

\subsubsection{Photon Trajectory}

Let us consider an initial photon emanating from the fireball located at the center of the coordinate. We define the unit momentum vector of initial photons in the Cartesian coordinate as 
$\bm{\hat{k}_i}=(\sin\vartheta\cos\varphi,\,\sin\vartheta\sin\varphi,\,\cos\vartheta)$, where $\vartheta$ and $\varphi$ are colatitude and azimuthal angle of initial photons, respectively. Assuming isotropic emission, $\cos{\vartheta}$ is chosen to be a random number in the range $\left[-1,1\right]$, whereas $\varphi$ is uniformly distributed in the range $\left[0,2\pi\right]$. Under the dipole magnetic field (eq. [\ref{eq:B-field}]), the incoming photon angle in the OF with respect to the local magnetic field line, $\Theta_i$, is related to the initial photon colatitude, $\vartheta$, via
\begin{equation}
    \label{eq:Theta_i vs vartheta}
\cos\Theta_i=\bm{\hat{B}}\cdot\bm{\hat{k}_i}=\frac{2\cos\vartheta}{\sqrt{3\cos^2\vartheta+1}}.
\end{equation}
This allows us to determine the particle velocity (eq. [\ref{eq:velocity distribution}]) and ERF cyclotron energy  (eq. [\ref{eq:resonance condition}]) in resonance with the incoming photon. 
The general relativistic effect on the photon trajectory (i.e., the light bending due to the gravitational redshift) is neglected because the altitude of the scattering layer is usually high ($\gtrsim$ several stellar radii) enough to avoid this.

\subsubsection{Scattering Probability}
In general, the scattering probability depends on the local plasma density, and thus we need to assume the spatial current distribution to know whether the scattering occurs at the resonance point. This approach may be useful for the persistent emission from a magnetar in its
quiescent state, if the steady electric currents are induced along the twisted magnetic field lines in a similar manner to that of the pulsar force-free magnetosphere (e.g., \citealt{TLK2002,FT2007,Nobili2008}). However, during the flare, when a dense cloud of particles are anticipated to be newly supplied in the magnetosphere,
the presence of such persistent currents is not trivial and hence a self-consistent treatment of the magnetosphere is not possible anymore. 

In order to avoid these complications, we presume that the plasma is sufficiently optically-thick to the resonant scattering (but optically-thin to the non-resonant scattering), which is likely the case for bursting magnetospheres (see Appendix \ref{s:tau}), and that any seed photon is scattered only once with $100$\% probability at the resonance point (e.g., \citealt{Nobili2008}), which mimics a situation where the optical depth to the RCS is of order unity.
In reality, the photon may experience multiple scatterings depending on the local RCS optical depth. Yet, it seems reasonable to begin with the single scattering case and thus we leave the exploration of multiple scattering for future work (see discussion in \S \ref{s:Discussion} for the limitation of our model). 
These assumptions make the scattering process entirely independent of both the magnetic field strength and the density of the scattering charges, and thereby considerably reducing the complexity.

\subsubsection{Emission Angle}
\label{sss:Emission Angle}
We follow the prescription by \citet{FT2007} to determine the direction of scattered photons in the ERF. The differential cross section of RCS is proportional to
\begin{equation}
\frac{d\sigma_{\rm res}}{d(\cos\theta_f)\, d\Pi}\propto1+\cos^2\theta_f,
\end{equation}
where $\Pi$ is its azimuthal angle about the local magnetic field direction (for more complete expression that takes into account quantum effects, see \citealt{Gonthier2014}). 
Thus, the cumulative probability density of scattering into an angle $\le\cos\theta_f$ is
\begin{equation}
\label{eq:scattering probability}
p=\frac{1}{8}(\cos^3\theta_f+3\cos\theta_f+4),
\end{equation}
which can be solved for $\cos\theta_{f}$ analytically:
\begin{equation}
\cos\theta_{f}=\left\{q+\sqrt{q^2+1}\right\}^{-1/3}-\left\{q+\sqrt{q^2+1}\right\}^{1/3},
\end{equation}
where $q\equiv2-4p$.
The distribution of $\cos\theta_{f}$ is then uniquely determined by randomly generating $p$ in the range $[0,1]$. 
Regarding the azimuthal angle about local magnetic field, we randomly choose $\Pi$ in the range $[0,2\pi]$ such that $\Pi=0$ coincides with $\bm{\hat{B}}$-$\bm{\hat{z}}$ plane (see Figure  \ref{fig:schematic}).

\section{Simulation}
\label{s:Simulation}
We generate a large sample of seed photons ($N_{\rm MC}=10^7$) by Monte Carlo technique. For each photon, we assign the initial momentum and treat its resonant interaction with scattering particle probabilistically to obtain its post-scattering momentum in the OF. 
Let $\theta_k$ be the colatitude of the scattered photon in the OF such that $\cos\theta_k=\bm{\hat{k}_f}\cdot\bm{\hat{z}}$.
The observer viewing angle is defined as $\theta=\theta_v$, and photons with $|\theta_v-\theta_k|<\theta_{\rm beam}$ are sampled to obtain the reprocessed spectra, where $\theta_{\rm beam}$ is the finite angular width that is centered on the observer orientation (we set $\theta_{\rm beam}=1^{\circ}$). Additionally, we extract $\theta_v$-integrated spectra by collecting all the scattered photons regardless of their directions.
Since the scattered photon direction is axsymmetric about $z$-axis (magnetic axis), there is no phase (i.e., $\phi$-direction) dependence for an aligned rotator. 

We note that the Lorentz factor of scattering particles in the magnetic polar region ($\Theta_i \sim 0$ or $\theta_v\sim0$) should be so high (see eq.[\ref{eq:velocity distribution}]) that the recoil effect becomes increasingly important. Since such an effect is neglected in our scattering treatment, the spectrum viewed from the observer in near polar direction may not be physically meaningful. Thus, we only select viewing angles sufficiently large  $\theta_v\gtrsim10^{\circ}$ enough to avoid this when presenting the direction-dependent spectra. Meanwhile, we include small viewing angles when obtaining the angle integrated spectra because the possible contribution of scattered photons from the polar region to the total spectrum is negligibly small.

\begin{figure}
 \includegraphics[width=\columnwidth]{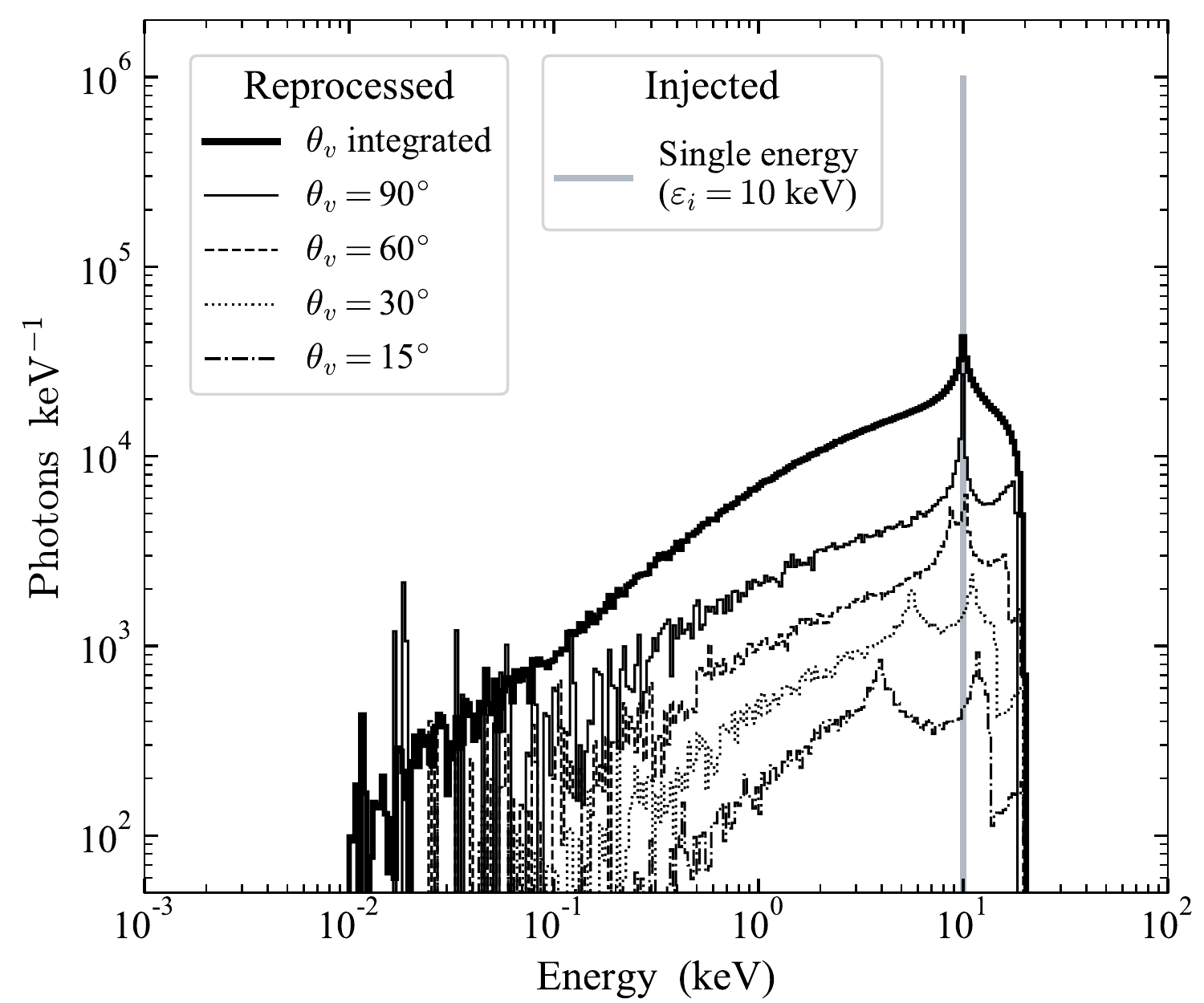}
 \caption{RCS spectra sampled for seed photons with energy $\epsilon_i=10$ keV (thick grey solid histogram) which is typical energy of thermal X-rays emanating from the fireball photosphere. The thick black solid histogram shows the reprocessed spectrum for photons averaged over the viewing angles. The thin black histograms (artificially scaled for demonstration purposes) correspond to spectra of photons that fall into different viewing angles. 
}
 \label{fig:dNdE_monoenergetic}
\end{figure}

Figure \ref{fig:dNdE_monoenergetic} illustrates the transmitted spectra for seed photons with a single energy $\epsilon_i=10$ keV. Evidently, one can see the effect of both up-scattering ($\epsilon_f>\epsilon_i$) and downscattering ($\epsilon_f<\epsilon_i$).
Remarkably, the maximum degree of up-scattering
$\epsilon_f/\epsilon_i\sim2$ is modest albeit fully consistent with the range that our model limits: 
\begin{equation}
    0\le\epsilon_f/\epsilon_i=1+\cos\Theta_i\cos\theta_f\le2,
\end{equation}
where equations \eqref{eq:resonance condition}, \eqref{eq:velocity distribution} and \eqref{eq: epsilon_f} are combined. This can be qualitatively understood as follows. Both (up- and down-) scatterings are pronounced when $|\cos\Theta_i\cos\theta_f|\sim1$. If one considers, for example, a case of $\cos\Theta_i\sim1$ and $\cos\theta_f\sim\pm1$, this indicates a relativistic velocity of the scattering particle $\beta_e=\cos\Theta_i\sim1$ (see eq. [\ref{eq:velocity distribution}]) in our model. Such an electron scatters the photon in a parallel direction to the local magnetic field in the OF (i.e., $\cos\Theta_f\to1$ as $\beta_e\to1$ in eq. [\ref{eq:Doppler trans 2}]), which is independent of the ERF emission angle $\theta_f$. Meanwhile, $\cos\theta_f\sim\pm1$ can be realized with a certain probability since $\cos\theta_f$ is uniformly sampled from the range $[-1,1]$. Therefore, the strong (up- and down-) scattering should be observed in the near polar directions (i.e., the bimodal energy redistribution seen in the $\theta_v=15^{\circ}$ case), whereas in the near equatorial directions (see the $\theta_v=90^{\circ}$ case; there is little energy redistribution) scattering should be relatively suppressed, which accounts for the substantial differences among spectra viewed by the observer in different directions. 

\begin{figure}
 \includegraphics[width=\columnwidth]{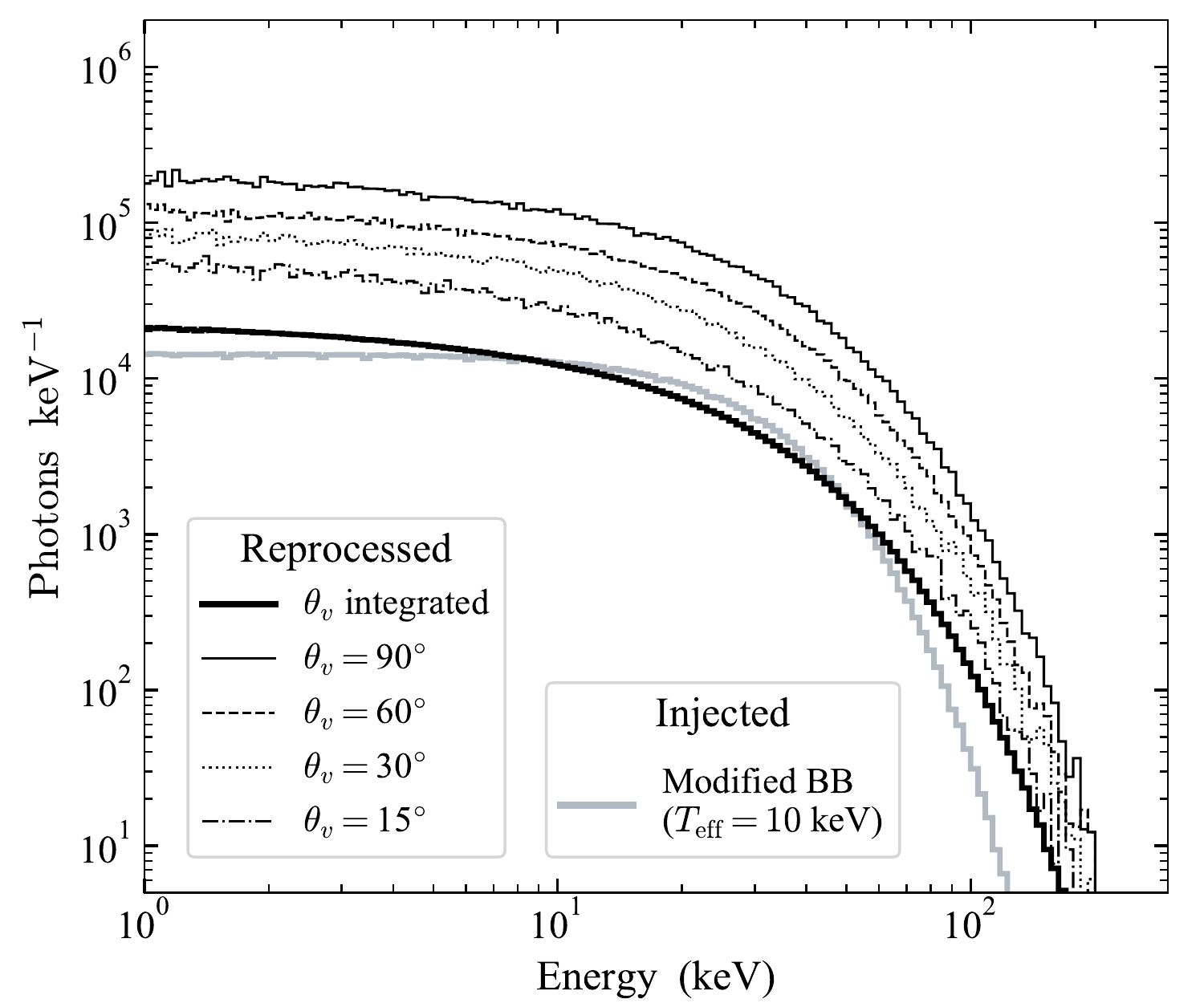}
 \caption{RCS spectra that might be sampled during magnetar flares (black histograms). Thin black histograms are artificially scaled for demonstration purposes.
 The seed photon spectrum (the modified blackbody spectrum proposed by \citealt{Lyubarsky2002}; see \S \ref{sss:Seed Photon Spectrum}) with an effective temperature of $T_{\rm eff}= 10$ keV is also shown by the thick grey histogram.
}
 \label{fig:dNdE_broadband}
\end{figure}

In Figure \ref{fig:dNdE_broadband}, we contrast the reprocessed broadband spectra with the injected fireball spectrum given by equation (\ref{eq:fireball spectrum}) with an effective temperature of $T_{\rm eff}=10$ keV. One can see that the injected fireball spectrum is Compton up-scattered by a factor of $\sim2$ at $\epsilon\gtrsim30$ keV. Moreover, the lower energy spectrum also exhibits a noticeable change; the initially flat spectrum becomes somewhat steeper due to the down-scattering. Furthermore, despite the clear angular dependence of the scattering seen in the case of mono-energetic spectrum shown in Figure \ref{fig:dNdE_monoenergetic}, there is little difference among the reprocessed broadband spectra of the aligned rotator viewed in different angles, which clearly indicates that our model is almost isotropic.

\begin{figure}
 \includegraphics[width=\columnwidth]{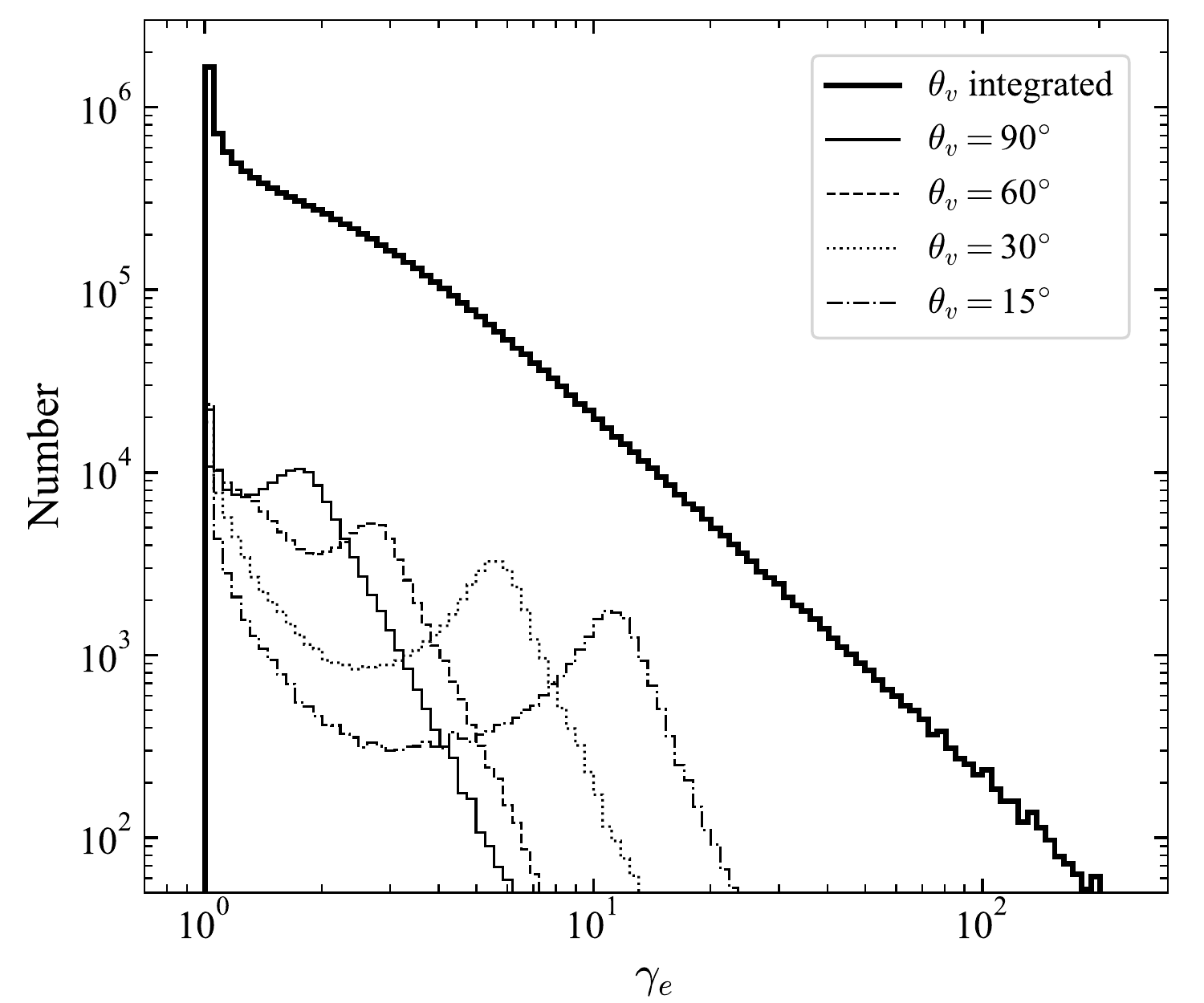}
 \caption{Lorentz factor distribution of electrons or positrons that scatter photons in different viewing angles (thin black histograms). The result for all scattering particles (integrated over the viewing angles) is indicated as a thick black solid histogram, which demonstrates that the majority of scattering particles move at the mildly-relativistic velocity.
}
 \label{fig:gamma_e_10keV}
\end{figure}

This is largely due to the mildly-relativistic energies of scattering particles in our simulation as shown in Figure \ref{fig:gamma_e_10keV}; relativistic Doppler boost is insignificant. 
The near-isotropic nature of our broadband model also suggests that any degree of misalignment between the spin axis and the magnetic moment (i.e., temporal variation in the effective viewing angle) would not produce any noticeable modulation in the reprocessed spectrum.
Accordingly, we adopt the $\theta_v$-integrated spectrum for the aligned rotator as a fiducial model. Thus, we can characterize the model with virtually only one free parameter $T_{\rm eff}$ (see \S \ref{ss:Fitting Procedure} for the fitting procedure).

\section{Application to Energetic Flares}
\label{s:Application}
In this section, we apply our model spectra
presented in \S \ref{s:Simulation} to the energetic flares from SGR 1900$+$14 which is one of the best observed magnetar with the surface dipole field strength $B_p\sim7.0\times10^{14}$ G and spin period $P\sim5.2$ s \citep{Kouveliotou1999,Hurley1999}. 
The results would be an important benchmark for applications to other sources.

\subsection{Fitting Procedure}
\label{ss:Fitting Procedure}
Since our model implementation is purely numerical, a formal fit to the data requires the interpolation among a grid of pre-calculated spectral templates.
We generate the spectral templates over a wide range of effective temperature $T_{\rm eff}$ in $1$--$40$ keV with a uniform grid spacing of 1 keV 
for a fixed array of energy bins $\epsilon$, which are assumed to be contiguous. In this work, we set $\epsilon/({\rm keV})\in[1,~300]$ such that the spectrum is well-sampled over the energy range of interest, which depends on the data/detector in question. For given values of $T_{\rm eff}$ and $\epsilon$, the model is calculated by linear interpolation of the $T_{\rm eff}$ and $\epsilon$ grid.

As demonstrated in \S\ref{s:Simulation}, our model provides the spectral photon counts, $N=N(T_{\rm eff};\,\epsilon)$ in units of photons${\rm \ keV^{-1}}$ (see Figure \ref{fig:dNdE_monoenergetic} and \ref{fig:dNdE_broadband}), which linearly scales with the number of artificially injected photons in simulation ($N_{\rm MC}$; \S \ref{s:Simulation}).
Thus, we define $\widetilde{N}(T_{\rm eff};\,\epsilon)=N/N_{\rm MC}$ in units of scattered photons${\rm\ keV^{-1}}$ per incident fireball photon.
It can then be integrated over the energy range of the detector to calculate the total number of scattered photons per incident fireball photon within that range, $\widetilde{n}_{\rm det}=\widetilde{n}_{\rm det}(T_{\rm eff})$ (some photons may fall outside of the detector's
energy range, causing this quantity to be less than unity). Then, the expected spectral photon flux is explicitly described as
\begin{equation}
\label{}
N_{\rm model}(T_{\rm eff};\,\epsilon)=\widetilde{N}(T_{\rm eff};\,\epsilon)\times \frac{N_{\rm obs}}{\widetilde{n}_{\rm det}(T_{\rm eff})},
\end{equation}
where $N_{\rm obs}$ is the measured number of photons in units of photons ${\rm cm^{-2}\ s^{-1}}$ within the relevant energy range of the detector. 

Finally, the parameter $T_{\rm eff}$ is determined by fitting $N_{\rm model}$ to $N_{\rm data}$ using Markov Chain Monte Carlo (MCMC) sampling\footnote{
If one only has a single free parameter, a Monte Calro simulation might be sufficient. Nevertheless, here we use a MCMC sampler in mind of possible implementation of our model with additional parameters in the future.} with {\fontfamily{qcr}\selectfont emcee}, a Python based affine-invariant sampler \citep{Foreman-Mackey2013}.  We define the likelihood function as $\ln\left({\cal L}\right)=-\chi^2/2$, and adopt a uniform prior for the temperature in $T_{\rm eff}/({\rm keV})\in \mathcal{U}(1,40)$. 
We obtained the best-fit parameter using $400$ walkers and $3,000$ steps ($\sim10^6$ total samples). The samplers were initialized in a small Gaussian sphere enclosing the preferred model parameter, after some iteration. The assessment of the model fitting is given in Appendix \ref{s:Model Fitting}.

\subsection{Results}
\subsubsection{Intermediate Flares}
\label{sss:IF}

\begin{figure}
 \includegraphics[scale=0.84]{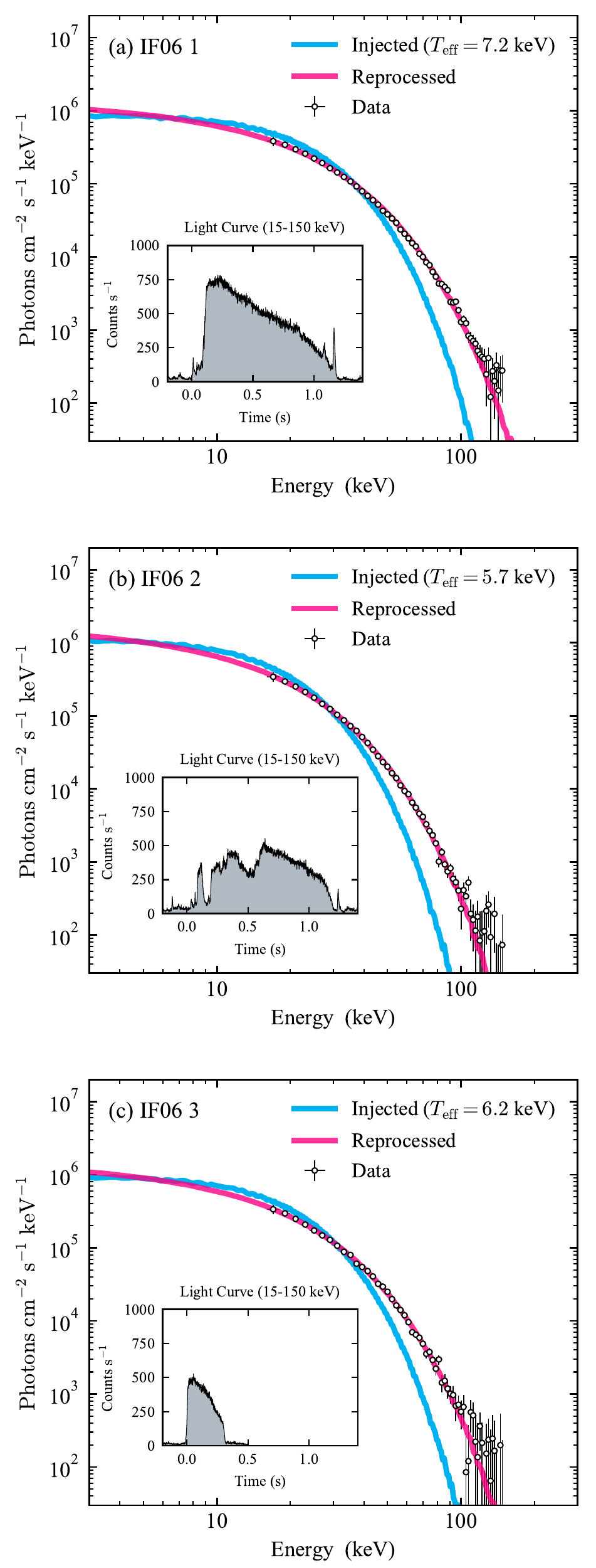}
 \caption{Hard X-ray spectra of the intermediate flares from SGR 1900$+$14 observed on 2006 March 29 by {\it Swift}/BAT. The data are fitted by the reprocessed model spectra (red; \S \ref{s:Simulation}). The injected model spectra (the modified blackbody spectrum proposed by \citealt{Lyubarsky2002}) are also shown (blue; \S \ref{sss:Seed Photon Spectrum}). Insets show flare light curves.
}
 \label{fig:IF06}
\end{figure}

\begin{table*}
\smallskip 
\begin{center}
\caption{SGR 1900$+$14 burst properties on March 29, 2006 (IF06) and August 27, 1998 (GF98). The effective fireball temperatures ($T_{\rm eff}$) derived from the best fits to the individual flare spectrum are also shown with errors of 1-$\sigma$. The reduced-$\chi^2$ values corresponding to the median likelihood are also presented
to indicate the goodness of fit.}
\label{tab:SGR1900+14}
\begin{threeparttable}
\begin{tabular}{ccccccccc}
\cmidrule[0.8pt](r){1-9}
  Burst &Energy Range &Time & Duration & Luminosity\tnote{a} & Energy\tnote{a}   & Fireball Temperature $T_{\rm eff}$ & $\chi^2/{\rm dof}$&\\
 &(keV)&(UT)& (s) & (${\rm erg\ s^{-1}}$) & (erg) & (keV) & &\vspace{1mm}  \\
  \cmidrule[0.4pt](r){1-9}
 IF06 1 &\multirow{3}{*}{$15$--$100$}&$02$:$53$:$13.3$&$1.2$ & $1.5\times10^{41}$ & $1.8\times10^{41}$ &  
 $7.23^{+0.03}_{-0.03}$ & $52/57=0.91$ &\\
 IF06 2 &&$02$:$53$:$15.4$& $1.2$ & $1.0\times10^{41}$ & $1.2\times10^{41}$&  
 $5.71^{+0.03}_{-0.03}$ &  
 $40/57 = 0.71$ &\\
 IF06 3 &&$02$:$53$:$22.3$&$0.34$ & $1.1\times10^{41}$ & $3.7\times10^{40}$&  $6.15^{+0.04}_{-0.04}$ & $34/57 = 0.59$ &\\
 \cmidrule[0.4pt](r){1-9}
 GF98 B &\multirow{2}{*}{$40$--$700$}& $10$:$23$:$24.1$ & $128$ & $2.9\times10^{40}$ & $3.7\times10^{42}$ &$32.81^{+0.12}_{-0.15}$ &$ 3908/71 = 55.04$ &\\
 GF98 C && $10$:$25$:$32.1$ & $128$ & $4.6\times10^{39}$ & $5.9\times10^{41}$ &   $29.41^{+0.80}_{-0.62}$ & $ 154/39 = 3.96$&\\
 \cmidrule[0.8pt](r){1-9}
\end{tabular}
\begin{tablenotes}
\item[a] Assuming a distance to the source of $10$ kpc (e.g., \citealt{Olive2004,Israel2008}).
\end{tablenotes}
\end{threeparttable}
\end{center}
\end{table*}
We first analyze intermediate flares, which occurred on 2006 March 29, using the data of Neil Gehrels {\it Swift} Burst Alert Telescope (BAT; \citealt{Krimm2013}).
Following the same methodology as presented in \citet{Israel2008}, we carry out spectroscopy in the $15$--$150$ keV range\footnote{Since the cross section of photoelectric absorption drops exponentially at hard X-ray energies, the extracted spectra are not affected by the interstellar absorption; they can be regarded as intrinsic burst spectra.}. Specifically, we extract time-integrated spectra of three intermediate flares (IF06 1--3 hereafter; see Table \ref{tab:SGR1900+14}), occurring at $4.0$--$5.2$ s, $6.1$--$7.3$ s and $13$--$13.3$ s in the top panel of Figure 1 in \citet{Israel2008}, respectively. They are all classified as intermediate in terms of duration and luminosity, albeit close to the lower end of the criteria. 
In such high-luminosity flares, the relaxation timescale for particle motion $\tau_{\rm relax}\sim{\cal O}(10^{-8}\,{\rm s})\,L_{40}^{-1/2}$ (see eq. [\ref{eq:t_relax}] in Appendix A) is sufficiently shorter than the minimum propagation timescale $\sim R_{\rm NS}/c\sim{\cal O}(10^{-5}\,{\rm s})$, and therefore the self-consistent particle velocity field (eq. [\ref{eq:velocity distribution}]) should be always maintained, which makes them ideal targets for our model.
We summarize our fitting results in Table \ref{tab:SGR1900+14} and compare our best-fit models with the observed burst spectra in Figure \ref{fig:IF06}.
As one can clearly see, our model shows surprisingly good agreement with observations, yielding $\chi^2/{\rm dof}$ values near unity.
We obtain best-fit effective temperatures $T_{\rm eff}=6$--$7$ keV for these bursts.

Although the soft-band X-ray spectra are unavailable for these bursts, since our model predicts that the downscattering effect would slightly steepen the initially flat spectrum at low energy ranges, it is also interesting to examine this with observations. 
\citet{Olive2004} observed another set of intermediate flares with average luminosity of $L=6.0\times10^{40}\ {\rm erg\ s^{-1}}$ from the same source that occurred on 2001 July 2, using the data of FREGATE (French Gamma-Ray Telescope) and WXM (Wide-Field X-Ray Monitor) experiments aboard the HETE (High-Energy Transient Explorer) spacecraft. They obtained a broadband spectrum over $2$--$150$ keV, extending down to soft X-ray energies (see Figure 9 of \citealt{Olive2004}). We confirm that our model appears in good agreement with their broadband spectral behaviour including the softer part. 

\subsubsection{Giant Flare Extended Tail}
\label{sss:GF}

\begin{figure}
 \includegraphics[width=\columnwidth]{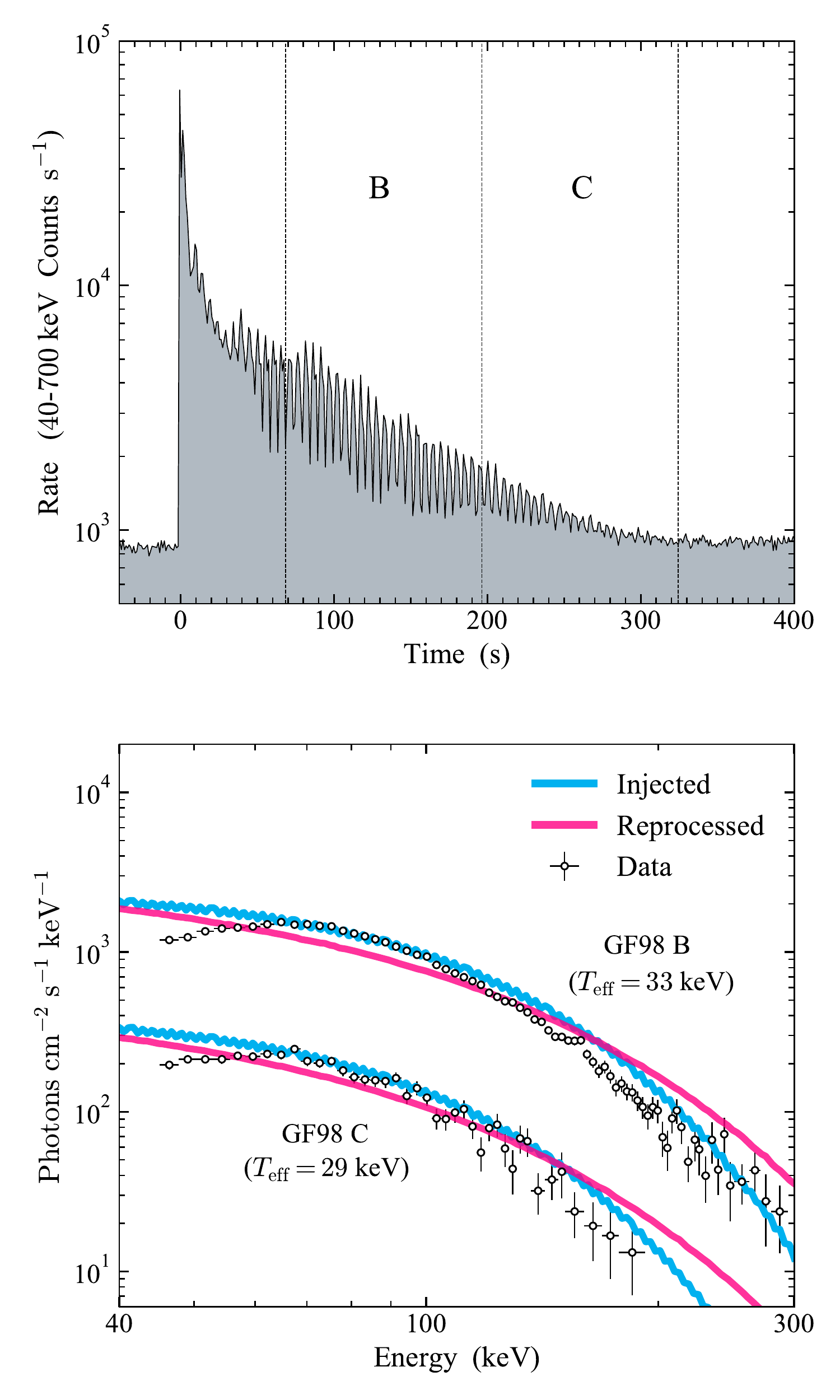}
 \caption{Hard X-ray to soft gamma-ray light curve (top) and spectra (bottom) of the giant flare extended tail from SGR 1900$+$14 observed on 1998 August 27 by the {\it BeppoSAX}/GRBM. The spectra are extracted from and averaged over two contigious 128-s intervals (denoted as B and C in the top panel) significantly after the initial spike ($68.4$ s and $196.4$ s after the flare onset, respectively). Similarly to Figure \ref{fig:IF06}, the data are fitted by the reprocessed model spectra (red; \S \ref{s:Simulation}) and their corresponding injected model spectra are also shown (blue; \S \ref{sss:Seed Photon Spectrum}).}
 \label{fig:GF98}
\end{figure}

Additionally, we analyze the historical giant flare which occurred on 1998 August 27 (hereafter GF98; \citealt{Hurley1999,Mazets1999,Feroci2001}), using the data of {\it BeppoSAX} Gamma-Ray Burst Monitor (GRBM; \citealt{Feroci1997}). Following the similar methodology as presented in \citet{Guidorzi2003}, we performed spectroscopy in the $40$--$700$ keV range. The GF98 comprises of two successive components. Initially, a short hard spike appears, subsequently followed by a gradually decaying tail which shows intense pulsation over $\sim300$ s (see the top panel of Figure \ref{fig:GF98}). The giant flare spectra during the early extended tail phase including the initial spike are described by a combination of thermal and power-law emission \citep{Hurley1999}. The pulsating thermal component is likely due to emission from the trapped fireball, whereas the hard non-pulsating component visible only at the early phase ($\lesssim40$ s) requires an additional explanation, such as emission from the heated corona around the trapped fireball \citep{TD2001}. We therefore select two successive 128-s intervals denoted as B and C at the sufficiently late phase of extended tail (see Table \ref{tab:SGR1900+14}) in order to obtain pure spectra of trapped-fireball origin. As summarized in Table \ref{tab:SGR1900+14}, we obtain best-fit effective temperatures $T_{\rm eff}\sim30$ keV with large $\chi^2/{\rm dof}$ values. Thus, unlike the case of intermediate flares, our model does not describe the observed giant flare spectra adequately, which is also seen in the bottom panel of Figure \ref{fig:GF98}.    

\section{Discussion}
\label{s:Discussion}

As shown in the previous section, the observed intermediate flare spectra at $15$--$100$ keV are successfully reproduced by our RCS model. The reduced-$\chi^2$ values for intermediate flares are slightly less than one ($\sim0.6$--$0.9$), which could indicate over-fitting or overestimated uncertainties.  In any case, the data suggests that a very simple model is a good choice. The best-fit effective temperatures of emission allow us to estimate the spherical radius of the trapped fireball $r_{\rm FB}$ via $L=4\pi r_{\rm FB}^2\sigma_{\rm SB}T_{\rm eff}^4 $, where $\sigma_{\rm SB}=\pi^2/(60\hbar^3c^2)$ is the Stefan-Boltzmann constant and the emission is assumed to be isotropic. Adopting typical intermediate burst properties, this is rephrased as $r_{\rm FB}/R_{\rm NS}\sim2.5\ L_{41}^{1/2}\,(T_{\rm eff}/{6\,\rm keV})^{-2}$ (see also \citealt{Olive2004} for a similar discussion), which is comparable to the stellar radius. Meanwhile, a typical height of resonant layer $r_{\rm res}$ for the thermal emission with $T_{\rm eff}=6\,$--$\,7\;$keV is estimated by equations \eqref{eq:resonance condition}, \eqref{eq:cyclotron energy} and \eqref{eq:velocity distribution} as $r_{\rm res}/R_{\rm NS}\sim10\,\gamma_e^{1/3}$ for $\epsilon_i/B_{p,14}\sim1\,$keV.
Accordingly, we can ensure that the point-like assumption of fireball ($r_{\rm FB}\ll r_{\rm res}$) is valid  for these bursts. The implication of the consistency between our model and observations is that a single scattering is sufficient to account for the hard component of observed spectra. This may also imply that the optical depth to the RCS might be of order unity in the magnetosphere during intermediate flares. 

Meanwhile, the inconsistency between our model and observed giant flare spectra may highlight a possible limitation of our toy model. Qualitatively, a possible explanation for the discrepancy could be that our single scattering scheme does not work for giant flares, even though the angular velocity distribution of the particles (eq. [\ref{eq:velocity distribution}]) is maintained. In our spectral analysis presented in \S \ref{s:Application}, the giant flare spectra are time-integrated over many spin cycles ($128 \ {\rm s}/P\sim25$ cycles), while the intermediate flare spectra are time-integrated over a tiny fraction of the spin period ($1\ {\rm s}/P\sim0.2$ cycles). In this respect, intermediate flares might indeed be a better test case for the formalism of single scattering in our model, as it allows us to see a snapshot of the magnetosphere. In contrast, in the giant flare spectra, we may see a lot of multiply up-scattered and/or down-scattered photons over the entire magnetosphere. Moreover, the extremely high plasma density in the magnetosphere naturally expected for giant flares should enhance the scattering rate, which may also support the multiple scattering picture. 
Another possibility may be the emission geometry; the fireball temperature for giant flares is so high that the fireball size could be comparable to the height of resonance points, which would break the assumption that seed photons are emitted in an isotropic manner from the point-like source.

Since the majority of magnetar flares have more or less thermal spectra, two blackbody models are known to provide successful fits.
These rather phenomenological models may be interpreted as a thermalized emission from E-mode and O-mode photospheres (e.g., \citealt{Israel2008,Kumar2010,Younes2014}), while it is not clear whether the observed difference in temperature and size between the two photospheres can be truly realized \citep{vanPutten2016}. 
In contrast to such models, the seed photon spectrum adopted in this work (\citealt{Lyubarsky2002}) takes into account the energy transfer under the presence of strong magnetic field and two photon polarization modes (as noted in \S \ref{sss:Seed Photon Spectrum}). Moreover, our model is a single-component, physically meaningful model; this is an advantage over the phenomenological multi-component models. 
Despite the good agreement
with theoretical predictions, we believe that it is essential to study more bursts from different sources, to definitely validate our interpretation of the data. 

\section{Conclusions}
\label{s:Conclusions}

In this work, we newly proposed a useful model for spectral modification of magnetar flares by considering the reprocess of the original fireball spectrum by the RCS in its simplest form. During the flare, photons emitted from the fireball should resonantly interact with the magnetospheric particles. We show by a simple thought experiment that such scattering particles are expected to move at mildly relativistic speed along closed magnetic field lines, which would slightly change the incident photon energy due to a Doppler shift. Based on this idea, we develop a toy (single scattering) model for the RCS during the flare and perform three dimensional Monte Carlo simulation by taking into account both the angular velocity distribution of particles that is unique to flaring magnetospheres and the realistic seed photon spectrum from the trapped fireball. We find that our spectral model is almost independent of the observer's viewing angle and can be captured by a single parameter; the effective temperature of the fireball, which greatly reduces the complications and allows us to fit the observed spectra with low computational cost. 

Our model is then applied to the data of energetic magnetar flares from SGR 1900+14. We show that our model gives a surprisingly good fit to intermediate flares. This implies that a single scattering is sufficient to account for the hard tails seen in observed spectra, and thereby suggesting that the optical depth to the RCS might be of order unity in the flaring magnetosphere.
On the other hand, giant flare extended tails cannot be fully explained by the current model alone. This may be because of longer duration (integration time) and denser plasma environment expected for those brightest flares, both of which may favor a multiple-scattering picture.
In order to identify what missing physics might reconcile these with observational constraints, we need a more realistic RCS model that includes multiple scattering scheme as well as the refined treatments of scattering cross section, polarization states of photons, etc., which we defer for the future work. For instance, the emission from a trapped fireball are polarization dependent (e.g., \citealt{Yang15,Taverna2017}), and RCS may impart polarization on the outgoing photons. Therefore, combining polarization with the RCS by particles with the equilibrium velocity field (considered in this work) would be one of interesting directions that could be potentially probed with future hard X-ray polarimeters.
Regardless of those possible improvements, as our single-component, physically-motivated model can extract the information of the fireball, this could lead to a convenient tool to investigate magnetar bursts.

\section*{Acknowledgements}

We thank Chryssa Kouveliotou for invaluable assistance, Cristiano Guidorzi for helping with the {\it BeppoSAX}/GRBM data, John Livingston for useful suggestions on model fitting and the anonymous referee for a number of valuable suggestions that have significantly improved the quality of the manuscript. SY deeply thanks members of Astrophysics and Cosmology Group at Physics Department of Ben-Gurion University for their
hospitality during his stay when this work was initiated. SY greatly appreciates comments provided by Ken Ebisawa, Mareki Honma, Kunihito Ioka, Shimpei Shibata, Toshikazu Shigeyama and Tomonori Totani. SY was supported by the Research Fellowship of Japan Society for the
Promotion of Science (JSPS, Grant No. {\rm 17J04010}). JG is supported by 
the ISF-NSFC joint research program (Grant No. 3296/19), funded by the Israeli Science Foundation and the National Natural Science Foundation of China.

\section*{Data and Code Availability}
The data underlying this article and the codes used for generating the spectral models are available at \url{https://github.com/shotaro-yamasaki/flarespec}. These codes will be updated as advanced with tailor-made fitting tools.

\addcontentsline{toc}{section}{Acknowledgements}



\bibliographystyle{mnras}
\input{draft.bbl}


\appendix

\section{Timescale for regulation of particle motion}
\label{s: tau_relax}

Here we provide an order-of-magnitude estimate for the timescale over which relaxation of particle motion to dynamical equilibrium takes place. Following \citet{TLK2002}, the radiative force $F_{\rm rad}$ exerted on an electron at a given distance $r$ from the stellar center in the ERF is 
\begin{equation}
\label{eq:F_rad}
F_{\rm rad}=\int d\omega\, \sigma_{\rm res}(\omega)\frac{L_{\omega}}{4\pi r^2 c}\sim \frac{\pi^2r_e}{\omega_B} \, \frac{L}{4\pi r^2},
\end{equation}
where $L_{\omega}$ is the spectral intensity of the radiation and $L=\int L_{\omega}\,d\omega\sim(\omega L_{\omega})_{\omega=\omega_B}$ is the pseudo-bolometric luminosity. The angular dependence is neglected in the second equation for simplicity. 
Assuming typical parameters for energetic magnetar flares, the radiation force (eq. [\ref{eq:F_rad}]) is estimated as
\begin{equation}
\label{eq:F_rad explicit}
F_{\rm rad}\sim{\cal O}(10^{-6})\ B_{p,14}^{-1}\,L_{40}\,(r/R_{\rm NS})\quad {\rm dyne}.
\end{equation}
Consider a deviation of the incident radiation angle in the ERF with respect to the local magnetic field from the equilibrium state $\theta_i\neq\pi/2$; then an equation of particle motion along the magnetic field line gives 
\begin{equation}
\label{eq:eom}
    \gamma_{e}m_e\,\dot{v}_e=F_{\rm rad}\,\cos\theta_i,
\end{equation}
where $\dot{v}_e$ is the acceleration/deceleration along $\bm{B}$ relative to the initial ERF.
Given the size of the magnetic loop comparable to the location of the electron $\sim r$, the relaxation timescale for electrons to acquire the equilibrium velocity distribution can be roughly estimated as 
\begin{equation}
\label{eq:t_relax}
\tau_{\rm relax}\simeq\sqrt{\frac{r}{|\dot{v}_e|}}\sim \frac{{\cal O}(10^{-8})}{|\cos\theta_i|^{1/2}}\ B_{p,14}^{1/2}\,L_{40}^{-1/2}\gamma_e^{1/2}\quad {\rm s},
\end{equation}
where $\gamma_e\sim{\cal O}(1)$ in our model.

Note that the above is a rough estimate that assumes a starting point far away from the local equilibrium point ($\theta\sim0$). If an electron starts from near the equilibrium point ($\theta_i\sim\pi/2$), one can show that the relaxation timescale becomes even shorter. Making use of $\dot{v}_e\sim c\dot{\beta}_e\sim c(\beta_e-\cos\Theta_i)/\tau_{\rm relax}$ and equation \eqref{eq:Doppler trans 1},  equation \eqref{eq:eom} reads
\begin{equation}
\label{eq:t_relax_2}
\tau_{\rm relax}\simeq\gamma_e\left(1-\beta_e\cos\Theta_i\right)\frac{m_ec}{F_{\rm rad}}\to\frac{m_ec}{\gamma_eF_{\rm rad}}\sim
{\cal O}(10^{-11})\quad {\rm s},
\end{equation}
where a limit $\cos\Theta_i\to\beta$ is considered in the second transformation. Namely, as an electron gets near the equilibrium point it approaches it with an exponential decay to of $m_ec/F_{\rm rad}$.

Hence the timescale for relaxation of particle motion due to the radiation drag force is typically much shorter than the duration of the flare.  
One can also see that the above estimate holds for any location inside magnetosphere because the dependence on $r$ vanishes.

\section{Optical Depths to Resonant/Non-Resonant Scattering}
\label{s:tau}
In this work, we assume that the magnetosphere is optically thick to resonant scattering but optically thin to non-resonant one. 
The ratio of optical depth between the resonant scattering (eq. [\ref{eq:sigma_res}]) and the non-resonant scattering is
\begin{equation}
\label{eq:sigma_res at resonance}
\frac{\sigma_{\rm res}}{\sigma_{\rm T}}\sim{\cal O}(10^{5})\left(\frac{\hbar\omega_{\rm B}}{{\rm keV}}\right)^{-1}.
\end{equation}
Therefore, for any plasma number density and spatial scale, the resonant scattering dominates the non-resonant one. 
The condition that the magnetosphere is optically thick to resonant scattering but optically thin to non-resonant one ($\tau_{\rm T}\ll1\lesssim\tau_{\rm res}$) can be rephrased by means of plasma density:
\begin{equation}
\label{eq:tau condition}
{\cal O}(10^{12})\left(\frac{\hbar\omega_{\rm B}}{{\rm keV}}\right)  \lesssim \left(\frac{n_e}{{\rm cm^{-3}}}\right) \left(\frac{r_{\rm res}}{10\,R_{\rm NS}}\right) \ll {\cal O}(10^{17}),
\end{equation}
where $n_e$ is the mean plasma number density at the resonance point, which typically lies around $r_{\rm res}\sim10\,R_{\rm NS}$ in our model. Although it is challenging to estimate the local plasma density inside the flaring magnetosphere, estimates \citep{Beloborodov2007,Beloborodov2013} show  that the plasma density in magnetar magnetospheres exceeds the Goldreich-Julian density \citep{GJ1969}:
\begin{equation}
\label{eq:n_GJ}
n_{\rm GJ}(r_{\rm res})=\frac{B}{ecP}\sim{\cal O}(10^{10}) \ B_{p,14}\,\left(\frac{P}{1\,\rm{s}}\right)^{-1} \left(\frac{r_{\rm res}}{10\,R_{\rm NS}}\right)^{-3}\ {\rm cm^{-3}}.
\end{equation}
One can see that this is already close to the characteristic plasma density $\sim{\cal O}(10^{12}) \ {\rm cm^{-3}}$ required for the single resonant scattering (i.e., the first inequality in eq. [\ref{eq:tau condition}]), which implies that our assumption of single scattering may not be so unreasonable. In the quiescent magnetosphere, the equation \eqref{eq:n_GJ} should be factored by the pair multiplicity parameter ${\cal M}=100$--$1000$ \citep{Beloborodov2013}, and hence the above condition is more readily satisfied. Of course, when there is an additional supply of plasma, it is naturally expected that $\tau_{\rm res}\gg1$ and the multiple resonant scattering may come into play.  

\section{Assessment of Model Fitting}
\label{s:Model Fitting}

In this appendix, we test the framework of our spectral fitting method used in \S\ref{ss:Fitting Procedure}.
For this purpose, we synthesize mock data as follows. Consider a set of energy bins $\epsilon_j$ and their corresponding spectral photon fluxes $N_{\rm model}(\epsilon_j;\, T_{\rm eff}^{\rm true})$ with test parameter value $T_{\rm eff}^{\rm true}$. 
For each triplet $(\epsilon_j$, $N_{\rm model}(\epsilon_j)$, $\Delta N_{\rm model}(\epsilon_j))$, where $\Delta N_{\rm model}(\epsilon_j)$ denotes the error in the spectral photon flux of the mock data, we replace the second entry with a new value that is picked from a normal distribution around its model value $N_{\rm model}(\epsilon_j)$ with standard deviation given by a pure Poisson noise $\Delta N(\epsilon_j)=\sqrt{N(\epsilon_j)}$, i.e., 
$\tilde{N}(\epsilon_j)\sim{\cal N}(N_{\rm model}(\epsilon_j),\Delta N(\epsilon_j))$.
We choose the same set of energy bins spanning over $15$--$150$ keV as used in the real {\it Swift}/BAT data of intermediate flares (see \S\ref{sss:IF}).
The test parameter set for the mock data is chosen to be $T_{\rm eff}^{\rm true}/({\rm keV})\in\{5,~15,~25,~35\}$. 
With the mock data in hand, the first thing to do is to confirm that the data is consistent with the underlying test parameter values from where it was generated. We fit $N_{\rm model}$ to the synthetic data using MCMC and find that the marginalized best fit values recover the ``true" underlying parameter used to create the data set within the 1-$\sigma$ errors. We also confirm that the fitting results are insensitive to the error realizations and there is no apparent multi-modality in the likelihood distribution. Therefore, we ensure that our fitting method works correctly.


\bsp	
\label{lastpage}
\end{document}